\documentclass[preprint,12pt,authoryear]{elsarticle}

\usepackage[T1]{fontenc}
\usepackage[utf8]{inputenc}
\usepackage{lmodern}            
\usepackage{amsmath,amssymb,amsthm,bm,dsfont,mathtools}
\usepackage{graphicx}
\usepackage{booktabs,multirow,array}
\usepackage{nicefrac}
\usepackage{xcolor}
\usepackage{url}
\usepackage{hyperref}
\usepackage{appendix}
\usepackage[nameinlink,noabbrev]{cleveref}
\usepackage{subcaption}         
\usepackage{rotating}
\usepackage{enumitem}
\usepackage{float}
\usepackage{setspace}           

\begin{document}

\begin{frontmatter}

\title{The systemic impact of edges in financial networks}

\author[inst1,inst2,inst3]{Michel Alexandre\corref{cor1}}
\ead{michel.alsilva@gmail.com}

\author[inst1,inst4]{Thiago Christiano Silva}

\author[inst5]{Francisco A. Rodrigues}

\cortext[cor1]{Corresponding author.}

\affiliation[inst1]{organization={Research Department, Central Bank of Brazil},
            addressline={SBS Quadra 3 Bloco B},
            city={Brasília}, postcode={70074-900}, country={Brazil}}
\affiliation[inst2]{organization={Brazilian Institute of Education, Development and Research (IDP)}, country={Brazil}}
\affiliation[inst3]{organization={Department of Economics, University of São Paulo}, country={Brazil}}
\affiliation[inst4]{organization={Universidade Católica de Brasília},
            city={Brasília}, country={Brazil}}
\affiliation[inst5]{organization={Institute of Mathematics and Computer Science,\\ University of São Paulo},
            city={São Carlos, SP}, country={Brazil}}

\begin{abstract}
In this paper, we assess how the stability of financial networks is affected by interconnectedness considering its tiniest variation: the edge. We compute the impact of edges as the percentage difference in the systemic risk (SR) of the whole network caused by the inclusion of that edge. We apply this framework to a thorough Brazilian dataset to compute the impact of bank-firm edges. After observing that (i) edges are heterogeneous regarding their impact on the SR, and (ii) the fraction of edges whose impact on the SR is non-positive increases with the level of the initial shock, we use machine learning techniques to try to predict two variables: the criticality of the edges (defining as critical an edge whose impact on SR is significantly greater than that of the others) and the sign of the edge impact. The level of accuracy obtained in these prediction exercises was very high. These results have important implications for the development macroprudential policies aimed at financial stability. Our framework allows to identify, based on features related to the origin and destination nodes of the edge (i.e., the lending bank and the borrowing firm), whether an additional loan will have a significant and positive impact on the SR. 
\end{abstract}


\end{frontmatter}


\section{Introduction}
\label{sec:intro}
This paper deals with the impacts of interconnectedness on the stability of financial networks. A network can be represented as a collection of nodes (or vertices) connected through edges. Financial networks are ubiquitous, such as interbank markets \citep{allen2000financial,bardoscia2016distress}, bipartite bank-firm credit networks \citep{de2012bank,lux2016model}, networks of correlated assets \citep{mantegna1995scaling,gabaix2003theory}, and so on. The type of node varies according to the network (e.g., banks in interbank networks, banks and firms in bank-firm credit networks), and so does the type of edge (e.g., loans in bipartite bank-firm credit networks, return correlations in networks of correlated assets).

Mainly since the 2008 financial turmoil, network models have been widely applied to assess the relationship between interconnectedness and stability in financial networks \citep{eisenberg2001systemic,deng2021model,chen2024contagion,agosto2020tree,gai2010contagion,acemoglu2015systemic,glasserman2015likely,glasserman2016contagion,battiston2018financial,martinez2019interconnectedness}. More interconnections among different financial agents can promote shock propagation, leading to the collapse of a non-negligible part of the whole system. This risk is called \textit{systemic risk} (SR). On the other hand, interconnections are also a channel for risk-sharing. This dual role played by financial networks configures the so-called \textit{robust-yet-fragile} nature of financial networks \citep{chinazzi2015systemic}. Yet there is a not a consensus on the relationship between between interconnectedness and stability in financial networks, it is widely accepted that this relationship is nonlinear \citep{nier2007network,gai2010contagion}, and driven by other elements, such as the size of the initial shock \citep{acemoglu2015systemic} and the loss distribution regime adopted by distressed debtors \citep{alexandre2023does}.

In this paper, we assess how financial stability is affected by interconnectedness considering its tiniest variation: the edge. We compute the impact of edges as the percentage difference in the SR of the whole network caused by the inclusion of that edge. The SR of the financial network is computed through the \textit{differential DebtRank} (DDR) approach \citep{bardoscia2015debtrank}. Under this methodology, the SR corresponds to the loss of equity suffered by the financial system caused by the propagation of an exogenous initial shock, in the form of firms' equity loss (more on this on Section \ref{sec:meth}). For the computation of the SR, we consider different levels of the initial shock.

We apply this framework to a thorough Brazilian dataset to compute the impact on the SR of bank-firm edges -- that is, edges whose origin nodes are banks and destination nodes are firms. In other words, we are assessing the impact of edges that represent loans extended by banks to firms. In our dataset, the two types of agents -- banks and firms -- are connected through two financial networks: the interbank network and the bank-firm loan network. Two important results emerge from this preliminary analysis:

\begin{enumerate}
    \item Edges are quite heterogeneous concerning their impact on the SR of the financial network; and
    \item As far as the level of the initial shock increases, the fraction of edges whose impact on the systemic is null or negative increases.
\end{enumerate}

Concerning result (1), the distribution of the edges' impact is well represented by a log-normal distribution. This result is expected, as the most striking characteristic of complex networks -- the case of real financial networks -- regards the heterogeneity of their components \citep{albert2002statistical}. Therefore, some edges are \textit{critical} in the sense they impact the stability of the financial network much more than others. With respect to (2), the fraction of edges with null impact on the SR increases with the level of the initial shock. However, the fraction of edges with negative impact is much smaller and reaches a maximum for a certain level of the initial shock (more details on Section \ref{sec:res}).

In light of these two outcomes, we perform two prediction exercises. In both we employ machine learning (ML) techniques and a set of topological and financial features related to the origin and destination nodes of the edge (i.e., the lending bank and the borrowing firm) as predictive variables. First, we assess the drivers of the critical status of the edges. We define an edge as critical if it is an outlier according to the interquartile range (IQR) approach considering its impact. Thus, the critical edges are those whose inclusion would cause a large impact on the SR of the financial network. Shapley values are used in the interpretation of the results, showing which features are more relevant in the definition of critical edges in financial networks. We achieve a level of accuracy superior to 90\% in this prediction exercise.

Second, we try to predict the sign of the edge impact on the SR. The level of accuracy obtained in this prediction exercise was very high (96\%). Again we use Shapley values for the sake of a better interpretability of our results. This analysis shows the sign of the edges’ impact depends mostly on the PageRank (PR) of the edge’s origin node (the bank). More specifically, there is a critical PR $PR_{C}$ such that the probability of an edge originating from a bank with an PR lower than $PR_{C}$ having a non-increasing impact on systemic risk increases sharply.

Our paper contributes to at least two strands of the literature. The first one, as discussed previously, is related to the relationship between interconnectedness and SR in financial networks. The literature on this topic has already mentioned some drivers of this relationship, such as the shock size and the loss distribution regime. In this study, we argue that the impact on SR caused by a marginal variation in interconnectedness -- represented by the removal/inclusion of edges -- (i) is very heterogeneous in size, (ii) can be null or even negative in some situations, mainly when the size of the initial shock is large, and (iii) depends on some specific features of its origin and destination nodes.

The second one concerns the identification of critical edges. Although node importance has been explored more often,\footnote{See, for instance, \cite{alexandre2021drivers}, \cite{ghan2018}, \cite{kuzubas2014}, and \cite{jaramillo2014} specifically on the identification of important nodes in financial networks.} the study of the criticality of edges in networks has attracted much attention from researchers in many fields \citep{song2023important,yu2018identifying,de2020edge,brohl2019centrality,zhao2020identifying,hajarathaiah2024node}. Many approaches have been proposed to identify the importance of edges in networks \citep{onnela2007structure,yu2018identifying,helander2018gravity,zhao2020identifying,kanwar2022bc}. As far as we know, there is only one study \citep{seabrook2021evaluating} devoted specifically to the identification of critical edges in financial networks. The authors defined a structural importance metric, $l_{e}$, based on the change in the largest eigenvalues of the adjacency matrix of the network resulting from perturbations in it. The authors then propose a model of network evolution where this metric controls the probabilities of subsequent edge changes. Using synthetic data, they show how the parameters of the model are related to the capability of $l_{e}$ to predict whether an edge will change. Assessing five real networks (four financial and one social), the authors showed $l_{e}$ is slightly predictive of edge change in all cases, but only marginally so for two of the financial networks. The advantage of our approach is that it proposes a measure of edge importance explicitly related to its impact on the stability of the financial network.

Our results bring important implications in terms of macroprudential policy design. The withdrawal of edges is usually more feasible than the removal of nodes in financial networks as an strategy to promote financial stability \citep{zhao2020identifying}. By identifying which features of the lender and the borrower are mostly related to the magnitude and sign of the impact of the edge on the stability of the financial network, we show which edges should be removed or not included in order to promote financial stability. Our results suggest that by discouraging lending operations (through, for instance, capital buffers) between banks and firms with certain characteristics, policymakers may decrease the probability of the emergence of risk-increasing edges. Thus, we provide policymakers with a tool to identify the edges that should be targeted (i.e., edges with a positive and significant impact on financial stability) based on some characteristics of the lender and borrower.


This paper proceeds as follows. The dataset and methodological issues are discussed in Sections \ref{sec:data} and \ref{sec:meth}, respectively. Section \ref{sec:res} brings the results. Final considerations are presented in Section \ref{sec:concl}.

\section{The dataset}
\label{sec:data}

Our dataset comprises two types of information: i) financial and supervisory (F \& S) information on the agents, which can be financial institutions (FIs) or firms, and ii) information on the financial linkages between these agents. The FIs of our dataset can be banks or credit unions. All these data are from December 2022.


We include in our dataset FIs which are financial conglomerates or individual FIs with positive net worth. Concerning the firms, we include those belonging to the \textit{Alexandria} database \citep{docha2023industry} with positive net worth. The Alexandria database gathers information on more than 42,000 firms, of which more than 31,000 are non-financial companies (NFCs). Most of these firms are medium-sized and owned by non-residents. The firms belong to 18 different economic sectors, labeled from A to S (we are only considering NFCs, so financial firms are excluded). The F \& S information on these agents includes net worth, financial indicators (e.g., ROE), type of control, economic sector, and so on.

Regarding the financial linkages, we consider two financial networks: the interbank (IB) and the bank-firm credit network. In the IB network, we consider the net financial exposures between Brazilian FIs in the IB market. This network comprises all types of unsecured financial instruments, such as credit, capital, foreign exchange operations, and money markets. These financial instruments are registered in different custodian institutions -- Cetip, Central Bank of Brazil, and B3. Exposures between FIs belonging to the same financial conglomerate are excluded. The bank-firm credit network encompasses the corporate loans granted by FIs to non-financial firms. The source of this information is the Central Bank of Brazil’s Credit Risk Bureau System (SCR). Some topological metrics of both networks are presented in Table \ref{tab:topo}.

\begin{table}[t]
\begin{center}
\begin{tabular}{llr} \\
\hline
    Type of network & Metric & Value \\
    \hline \hline
    \multirow{6}{*}{IB network}&N. of nodes&803\\
    &N. of edges&2971\\
    &Assortativity&-0.3630\\
    &Average degree&7.3998\\
    &Average closeness centrality&0.1220\\
    &Average eigenvector centrality&0.0174\\
    \hline
    \multirow{6}{*}{Bank-firm network}&N. of nodes&11,955\\
    &N. of edges&28,157\\
    &Assortativity&-0.3965\\
    &Average degree&4.7105\\
    &Average closeness centrality&0.0002\\
    &Average eigenvector centrality&0.0063\\
    \hline
\end{tabular}
\end{center}
\caption{Some topological metrics of the networks}
\label{tab:topo}
\end{table}

\section{Methodology}
\label{sec:meth}

\begin{figure}[t]
    \centering
    \includegraphics[width=0.6\linewidth]{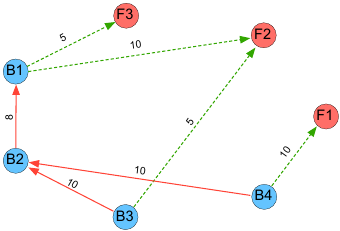}
    \caption{A stylized representation of a financial network}
    \label{fig:netw}
\end{figure}

As stated in Section \ref{sec:data}, we consider two types of financial linkages: the IB market and the bank-firm credit network. Figure \ref{fig:netw} depicts a simple representation of our financial networks. The nodes in blue are the banks, labeled as B1,...,B4, and the nodes in coral are firms, labeled as F1,...,F3. The edges in red represented the IB connections and those in green, the bank-firm linkages. Edges are directed, showing the origin node and the destination node, and weighted. For instance, bank 3 (B3) has lent an amount of resources equal to 10 to bank 2 (B2) in the IB market and equal to 5 to firm 2 (F2).  

The first step is to compute the SR of the whole system depicted in Figure \ref{fig:netw} by imposing losses in the form of net worth decrease to all firms at the same time. The net worth of both FIs and firms and the exposures among them in the two financial networks will be used at this step. The green edges are responsible for the \textit{direct losses} this initial shock will cause on the banks. The red edges represent the IB connections that will amplify this initial shock -- that is, they are responsible for the \textit{contagion}.

We are interested in measuring the systemic relevance of bank-firm edges -- that is, the green edges. This is done by removing one edge at a time and computing the proportional difference in the SR of the system. This metric is the edge's impact. An edge is deemed as critical if it is an outlier considering its impact, according to the IQR approach. Finally, we perform some prediction exercises. The topological variables built from the two networks, along with the F\& S variables on the agents and the variables related to the bank-firm loan itself, will constitute the set of potential explanatory variables. These variables will be employed to predict the target variables -- the edges' critical status and the sign of the edges' impact --- through ML techniques. In the following subsections, we detail our methodological process.

\subsection{Systemic risk}
\label{subsec:sr}

In what follows, we describe the \textit{differential DebtRank} (DDR) approach \citep{bardoscia2015debtrank}, used to compute the SR of the financial network. Let us define $E_{i}$ as the equity of agent $i$ and $A_{ij}$, the net exposure of agent \textit{i} towards agent \textit{j}. The agents correspond to banks and firms in Figure \ref{fig:netw}, and the exposures, to the weight of the edges between them. At period $t=0$, we impose an exogenous shock on the network, in which some agents will lose a fraction $\zeta$ of their equity. At $t=1$, the agents hit by the exogenous shock will transmit part of this loss to their creditors, by defaulting on part of their debt. In the next period, the agents that suffered some loss (that is, those with exposures towards the agents hit by the exogenous shock) will transmit part of this loss to their creditors. This process of loss transmission will continue in the subsequent periods. At a given period $t$, the accumulated loss transmitted by agent $j$ to its creditor $i$ is $L_{ij}(t)$. The aggregate loss suffered by $i$ (considering all $i$'s debtors) up to $t$ is $L_{i}(t)$. The dynamics of these two variables are represented by the following equations:

\begin{equation}\label{eq:loss}
    \Delta L_{ij}(t)=min\left(A_{ij}-L_{ij}(t-1),
    {A}_{ij}\frac{[L_{j}(t-1)-L_{j}(t-2)]}{E_{j}}\right),
\end{equation}

\begin{equation}\label{eq:aggl}
    \Delta L_{i}(t)=min\left(E_{i}-L_{i}(t-1),\sum_{j}\Delta L_{ij}(t)\right),
\end{equation}
in which $t\geq0$. In Eqs. \ref{eq:loss} and \ref{eq:aggl}, $\Delta L_{ij}(t)=L_{ij}(t)-L_{ij}(t-1)$ is the new flow of loss transmitted by \textit{j} to \textit{i} at $t$, and $\Delta L_{i}(t)=L_{i}(t)-L_{i}(t-1)$ is the variation in the total loss transmitted to \textit{i} by their debtors at \textit{t}. Thus, the loss transmitted by a given debtor to a given creditor is the proportional loss of equity suffered by the debtor times the exposure of the creditor on it. Observe, from Eq. \ref{eq:loss}, that the loss imposed by \textit{j} to \textit{i} cannot be greater than \textit{i}'s exposures towards \textit{j}. Moreover, an agent cannot suffer a loss greater than its equity (Eq. \ref{eq:aggl}).

At $t=T\gg0$, the system converges -- i.e., when no more losses are transmitted. Then, we compute the \textit{systemic risk} of the network according to the following equation:

\begin{equation}\label{eq:si}
    S_{\zeta}=100 \times \frac{\sum_{i}[L_{i}(T)-L_{i}(0)]}{\sum_{i}E_{i}}.
\end{equation}

Therefore, the systemic risk $S_{\zeta}$ of the network is the percentage of the aggregate equity of the system which is lost after an exogenous shock of size $\zeta$. Observe that we remove $L_{i}(0)$ from the computation, which is the loss resulting from the exogenous shock. Thus, $S_{\zeta}$ considers only the loss resulting from the contagion process.

\begin{figure}[t]
\centering
\begin{subfigure}{0.32\textwidth}
  \includegraphics[width=0.98\linewidth]{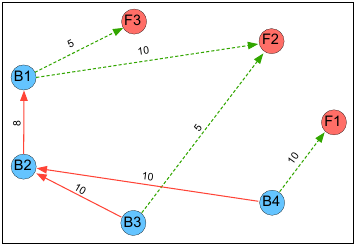}
  \caption{}
\end{subfigure}%
\begin{subfigure}{0.32\textwidth}
\centering
  \includegraphics[width=0.98\linewidth]{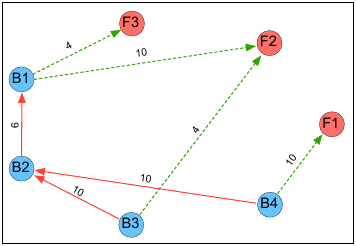}
  \caption{}
\end{subfigure}%
\begin{subfigure}{0.32\textwidth}
\centering
  \includegraphics[width=0.98\linewidth]{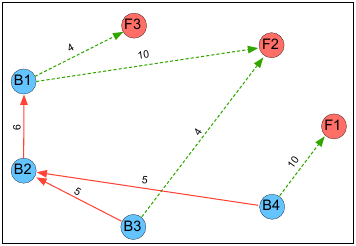}
  \caption{}
\end{subfigure}
\caption{Example of SR computation through the DDR approach}
\label{fig:ddr}
\end{figure}

Figure \ref{fig:ddr} gives a simple example of how SR is computed according to the DDR approach. Considering again Figure \ref{fig:netw}, suppose F3 loses 20\% of its net worth. This firm will default the same fraction on the exposures of B1, its only creditor, towards it. For simplicity, assume all banks have a net worth equal to 4. The loss transmitted by F3 to B1 ($0.2 \times 5 = 1$) corresponds to 25\% of B1's net worth. The exposure of B1 towards F3 reduces to 4 (Figure \ref{fig:ddr}, panel (a)). In the next  step, B1 defaults 25\% on the exposures of B2 in the IB towards it, imposing to B2 a loss ($0.25\times 8=2$) equal to 50\% of its net worth (Figure \ref{fig:ddr}, panel (b)). Finally, B2 defaults 50\% on its creditors' exposures, B3 and B4 (Figure \ref{fig:ddr}, panel (c)). However, as the loss suffered by an agent cannot be greater than its net worth (Equation \ref{eq:aggl}), this loss is capped at 4 for both B3 and B4. The process stops, as B3 and B4 have no creditors. The SR engendered by an initial shock of 20\% on F3's net worth is equal to the total loss excluding the initial shock ($2+4+4=10$) divided by the aggregate banks' net worth ($4\times4=16$) -- that is, 62.5\%.

\subsection{Classification of critical edges}
\label{subsec:class}

The next step is to set the critical status of the edges. For each edge $e$ of the network, we compute its impact in the systemic risk $I_{e,\zeta}$:

\begin{equation}
    I_{e,\zeta}=\frac{S_{\zeta}-S_{-e,\zeta}}{S_{-e,\zeta}},
\label{eq:impact}
\end{equation}
where $S_{-e,\zeta}$ is the systemic risk of the network for an exogenous shock of size $\zeta$ computed after the removal of edge $e$. If $I_{e,\zeta}>0$, it means the edge has a positive impact on the systemic risk of the network for that level of $\zeta$.

Considering the distribution of $I_{e,\zeta}$, we assign each edge a label $C_{e,\zeta}$ according to

\begin{equation} C_{e,\zeta} = 
\begin{cases}
    1,& \text{if } Q1-1.5IQR>I_{e,\zeta} \mid Q3+1.5IQR<I_{e,\zeta}\\
    0,              & \text{otherwise}
\end{cases}
\label{eq:crit}
\end{equation}    
where Q1 is the first quartile of the distribution of $I_{e,\zeta}$, Q3 is the third quartile, and $IQR$ is the interquartile range $Q3-Q1$. Therefore, an edge will be classified as critical if it is an outlier considering the distribution of $I_{e,\zeta}$, according to the IQR criterion. 

\subsection{Machine learning techniques}
\label{subsec:ml}
\begin{figure}[t]
    \centering
    \includegraphics[width=0.85\linewidth]{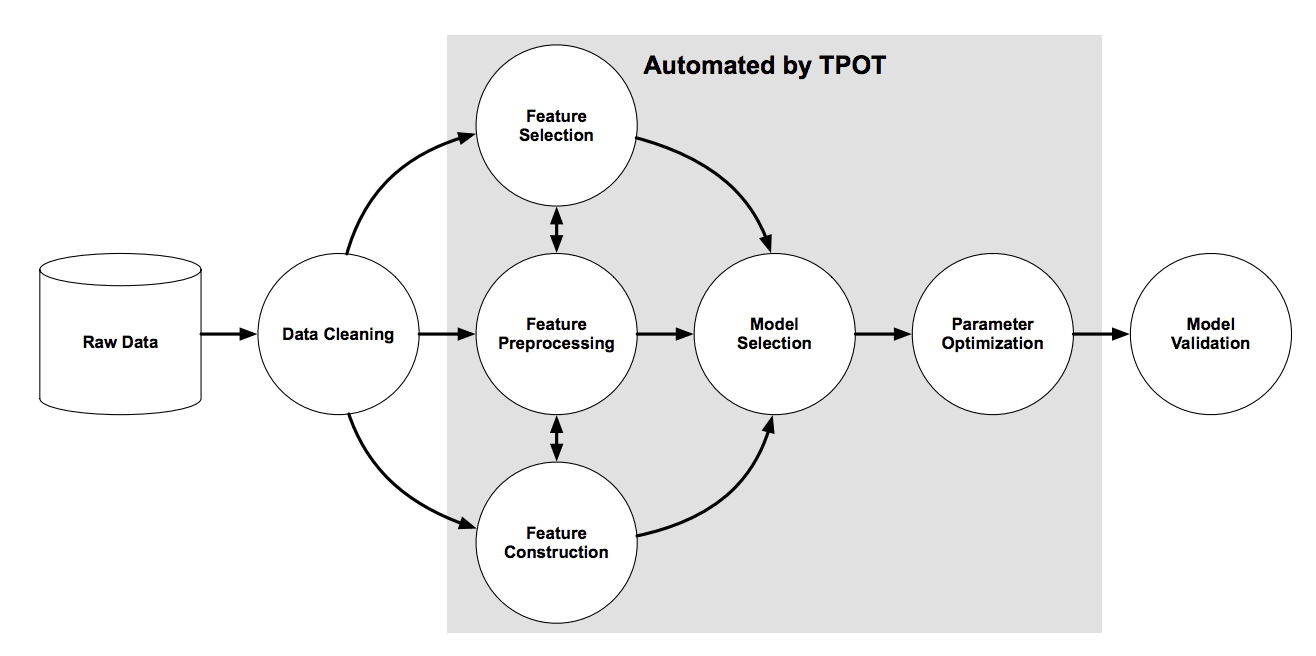}
     \caption{Overview of the TPOT pipeline search. Source: \cite{olson2016evaluation}}
    \label{fig:tpot}
\end{figure}

The final step is the prediction of the target variables -- the critical status of the edges and the sign of the edges' impact -- through ML techniques. For this task, we rely on the \textit{Tree-based Pipeline Optimization Tool} (TPOT), a programming-based automated machine learning (AutoML) system \citep{olson2016tpot}. TPOT uses a genetic programming (GP) stochastic global search procedure to efficiently discover the top-performing ML algorithm, as well as the optimal hyperparameters, for a given prediction problem. The search performed by TPOT encompasses all models included in the \textit{Scikit-learn} library, including linear (e.g., linear regression, logistic regression, ElasticNet, etc.) and nonlinear (SVM, tree-based models, XGBoost, neural networks, etc.) models. The TPOT pipeline was used to perform the following tasks: data cleaning, feature selection, feature processing, feature construction, model selection, hyperparameter optimization, and model validation (Figure \ref{fig:tpot}). The set of potential explanatory variables is presented in Table \ref{tab:vars}.\footnote{To ensure that the predictive variables are exogenous to the target variables, we exclude from this set features related to the edge itself, such as interest rate and maturity.}

For the interpretability of the predictions made by the ML model, we use the Shapley values approach, which originated from the coalition games theory \citep{shapley1953value,shoham2008multiagent}. This methodology informs on how important is a given feature in the prediction of the output, as well as whether this feature is positively or negatively correlated to the output. For the computation of Shapley values, we resort to the SHAP (SHapley Additive exPlanation) framework \citep{lundberg2017unified}. Let $g$ be an explainer model aiming at predicting an output. A set of $M$ features will be used as inputs. The predicted value for a given data instance is given by

\begin{equation}
    g(z')=\phi_{0}+\sum_{i=1}^{M} \phi_{i}z'_{i},
\end{equation}
where $\phi_{0}$ is the mean output, $\phi_{i}$ is the SHAP value of feature $i$, and $z'$ is a binary variable indicating whether feature $i$ was included in the model or not. Therefore, the SHAP value $\phi_{i}$ indicates to what extent the inclusion of feature $i$ in the model shifts (upwards or downwards) the predicted value from the mean output. If certain properties (local accuracy, missingness, and consistency) are met, $\phi_{i}$ corresponds to the original Shapley value \citep{lundberg2017unified}. The SHAP value of feature $i$ is defined by the following equation:

\begin{equation}
    \phi_{i}=\sum_{S \subseteq M\textbackslash i} \frac{|S|!(|M|-|S|-1)!}{M!} [F(S\cup \{i\})-F(S)].
\end{equation}

Therefore, the SHAP value of feature $i$ for a given data instance computes the difference between the predicted value of the instance using all features in $S$ plus feature $i$, $F(S\cup \{i\})$, and the prediction excluding feature $i$, $F(S)$. This is weighted and summed over all possible feature vector combinations of all possible subsets $S$. 

\section{Results}
\label{sec:res}
\subsection{General results}

We computed $I_{e,\zeta}$ (Eq. \ref{eq:impact}) for the bank-firm edges considering values of the initial shock $\zeta$ from 0.05 to 1.00, with step 0.05. The great majority of observations are positive. It means, in most of the cases, the inclusion of the edge increases the systemic risk. It corroborates the idea that financial networks are \textit{robust-yet-fragile} \citep{haldane2013rethinking}: if the financial network is sparse enough (what is the case of our network), shock propagation prevails over risk-sharing. Therefore, an increase in the interconnectedness of the network causes an increase in the systemic risk.

\begin{figure}[H]
    \centering
    \includegraphics[width=0.8\linewidth]{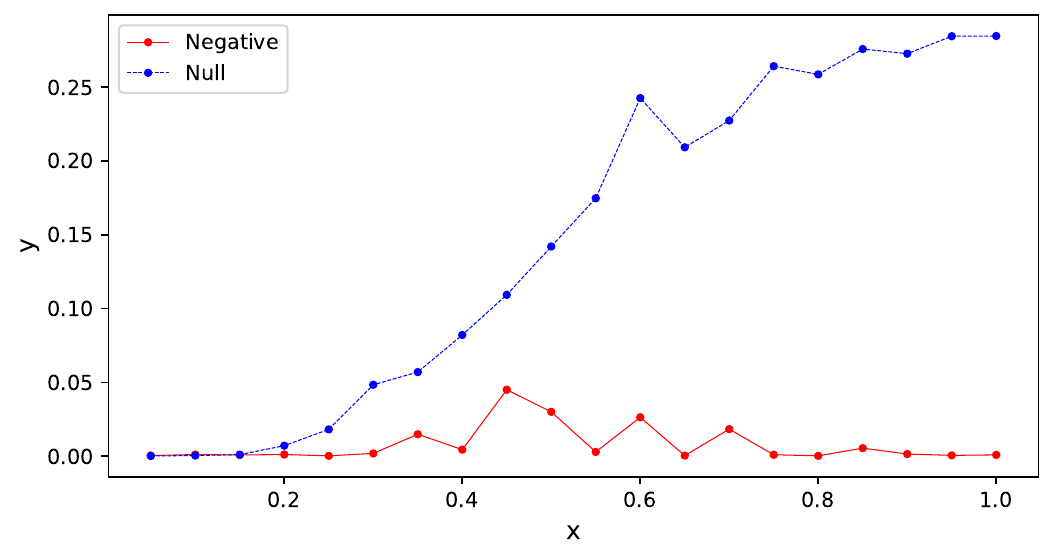}
    \caption{Fraction of edges with null or negative impact on the systemic risk of the financial network}
    \label{fig:edge}
\end{figure}

\begin{figure}[H]
    \centering
    \includegraphics[width=0.8\linewidth]{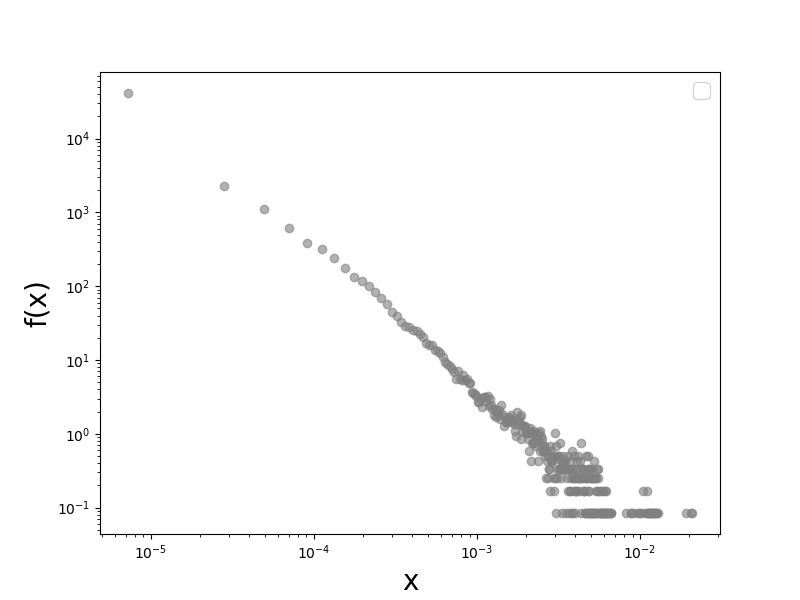}
    \caption{Distribution of edges' impact (in log scale)}
    \label{fig:distr}
\end{figure}

However, as can be seen in Figure \ref{fig:edge}, the fraction of edges with null impact on the systemic risk increases with the level of the initial shock. The fraction of edges with negative impact is much smaller and reaches its maximum for $\zeta\approx 0.45$. Moreover, Figure \ref{fig:distr} shows that the distribution of $I_{e,\zeta}$ is heavy-tailed. Comparing the goodness-of-fit of a power-law distribution to that of a log-normal distribution, one can see that the latter better fits the distribution of edges' impact.

\subsection{Predicting edges' critical status}
\label{ssec:pred_crit}
We then applied the TPOT approach in our set of predictive variables (Table \ref{tab:vars}) to predict the critical status of the bank-firm edges. We split our observations into in-sample (80\%) and out-of-sample (20\%) datasets. The in-sample dataset was oversampled through the Synthetic Minority Oversampling Technique (SMOTE) methodology \citep{chawla2002smote} so the fraction of critical edges reached 50\% before being used to train the model. We evaluate the model using accuracy as the performance metric through the \textit{k-fold cross-validation} technique, with $k=5$ and 10 repetitions. 

\begin{figure}[H]
    \centering
    \includegraphics[width=0.8\linewidth]{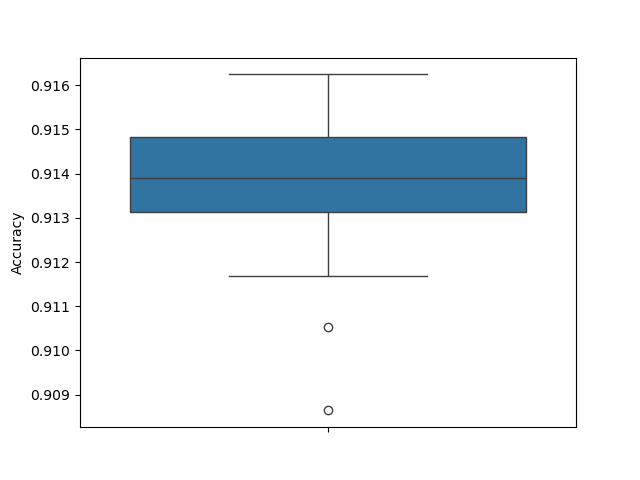}
    \caption{Histogram of accuracy -- in-sample dataset. Target variable: edges' critical status}
    \label{fig:acc_otl}
\end{figure}

The model selected by the TPOT pipeline was the \textit{XGBoost} \citep{friedman2000additive}. The optimal hyperparameters are presented in Table \ref{tab:hyper_otl}. In the in-sample dataset, the model has achieved an accuracy of over 90\% (Figure \ref{fig:acc_otl}). Table \ref{tab:perf_oos} presents some performance metrics for the out-of-sample dataset.

\begin{table}[H]
    \centering
    \begin{tabular}{llr}
    \hline
    Model & Hyperparameter & Value\\
    \hline \hline
        \multirow{5}{*}{XGBoost Classifier} &  learning\_rate & 1.0\\
         &  max\_depth & 4 \\
         &  min\_child\_weight & 1 \\
         &  n\_estimators & 100 \\
         &  subsample & 0.75 \\
    \hline
    \end{tabular}
    \caption{Hyperparameters of the ML models for the prediction of the edges' critical status}
    \label{tab:hyper_otl}
\end{table}

\begin{table}[H]
\centering
    \begin{tabular}{lcc}\\
    \hline
      Metric & Category = 0 & Category = 1\\
    \hline
    \hline
         Precision & 0.94 & 0.80 \\
         Recall & 0.95 & 0.78 \\
         F1-score & 0.95 & 0.79 \\
         Accuracy & \multicolumn{2}{c}{0.91} \\
         \hline
    \end{tabular}
    \caption{Performance metrics, out-of-sample dataset. Target variable: sign of edges' impact}
    \label{tab:perf_oos_otl}
\end{table}

\begin{figure}[t]
    \centering
    \includegraphics[width=0.8\linewidth]{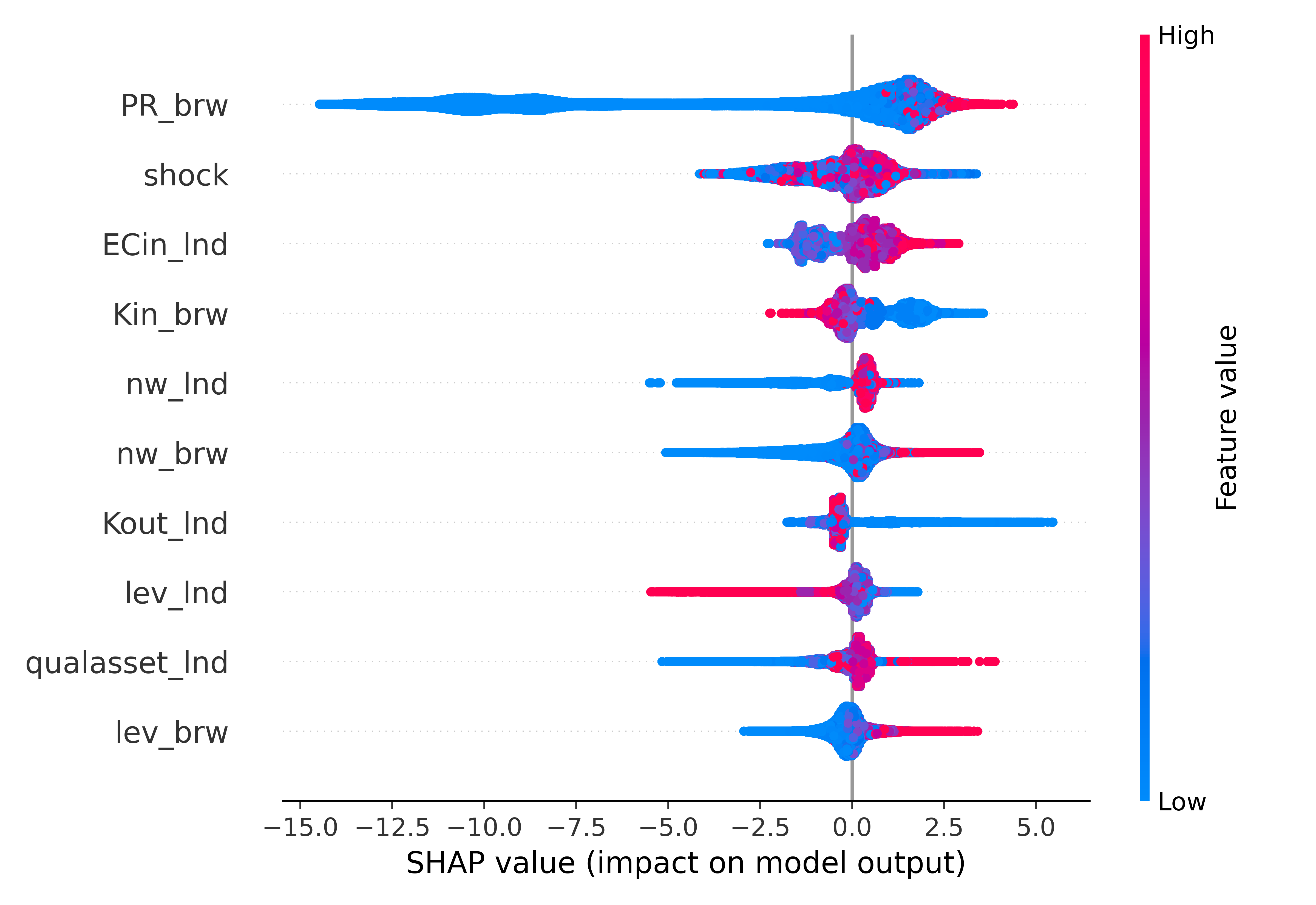}
    \caption{SHAP values, 10 most important features. Target variable: edges’ critical status}
    \label{fig:shap_otl}
\end{figure}

The SHAP analysis (Figure \ref{fig:shap_otl}) shows the main drivers of the critical status of the bank-firm edges. The figure can be interpreted as follows: for each feature, it is shown which observations have their predicted output increased (those at the right of the vertical axis) or decreased (those at the left of the vertical axis) after the inclusion of the feature in the model as a predictive variable. On the other hand, the observations closer to the red (blue) color spectrum have a higher (smaller) value of the feature. For instance, edges with a larger value of the borrower's PageRank (PR\_brw) are concentrated at the right side of the vertical axis -- that is, the probability of these edges being critical increases when this feature is used as a predictive variable. Thus, there is a positive correlation between the probability of an edge being critical and the value of the PageRank of the firm which is the destination node of this edge.

The features are sorted in descending order according to the mean absolute SHAP value. The relative importance of the features in predicting the critical status of the bank-firm edges varies according to the level of the initial shock. The PageRank of the borrower (that is, the firm) is the main driver of the edge's critical status, followed by the size of the shock and the eigenvector centrality (considering the incoming links) of the lender (ECin\_lnd). Both PR\_brw and ECin\_lnd are positively correlated to the probability of edges being critical. Thus, edges which are loans extended to firms with a large PageRank by banks with a large eigenvector centrality have a higher probability of being critical.

Figure \ref{fig:pdp_otl} depicts the SHAP partial dependence plots of the four most relevant features. These figures show how the expected value of the target variable varies according to the value of the feature. We can observe that some of these features present a nonlinear pattern of interaction with the target variable, corroborating the findings of Figure \ref{fig:shap_otl}. For instance, an increase in PR\_brw initially leads to a rise in the predicted value of the target for small values of the feature. However, if this value surpasses a given threshold, further increases in the value of the feature lead to a decrease in the predicted value of the target variable. 

\subsection{Predicting the sign of the edges' impact}

Next, we employed ML techniques to predict the sign of the edges' impact. The target variable is a categorical variable with value 1 if the $I_{e\zeta}$ is null or negative and zero otherwise. The set of potential explanatory variables is presented in Table \ref{tab:vars}. In order to guarantee a reasonable fraction of edges with non-positive impact, we considered only a value of $\zeta$ equal or greater than 0.5. As we did in the previous subsection, we split our observations into in-sample (80\%) and out-of-sample (20\%) datasets, the in-sample dataset was oversampled through the \textit{Synthetic Minority Oversampling Technique} (SMOTE) methodology, and the model was validated through the \textit{k-fold cross-validation} technique, with $k=5$ and 10 repetitions, using accuracy as the performance metric.

\begin{figure}[H]
    \centering
    \includegraphics[width=0.8\linewidth]{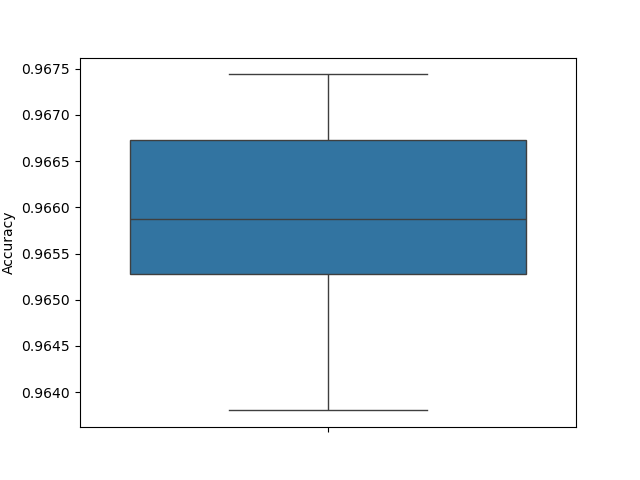}
    \caption{Histogram of accuracy -- in-sample dataset. Target variable: sign of edges' impact}
    \label{fig:acc_sign}
\end{figure}

The algorithm selected by the TPOT was the \textit{Decision Tree Classifier} \citep{safavian1991survey}. The optimal hyperparameters are presented in Table \ref{tab:hyper}. The accuracy obtained in the in-sample dataset was over 96\% (Figure \ref{fig:acc_sign}). The model also performed very well in the OOS dataset, as can be observed in Table \ref{tab:perf_oos}.

\begin{table}[H]
    \centering
    \begin{tabular}{llr}
    \hline
    Model & Hyperparameter & Value\\
    \hline \hline
        \multirow{4}{*}{Decision Tree Classifier} &  criterion & Gini\\
         &  max\_depth & 7 \\
         &  min\_samples\_leaf & 20 \\
         &  min\_samples\_split & 20 \\
    \hline
    \end{tabular}
    \caption{Hyperparameters of the ML models for the prediction of the sign of edges' impact}
    \label{tab:hyper}
\end{table}

\begin{table}[H]
\centering
    \begin{tabular}{lcc}\\
    \hline
      Metric & Category = 0 & Category = 1\\
    \hline
    \hline
         Precision & 0.98 & 0.93 \\
         Recall & 0.97 & 0.95 \\
         F1-score & 0.98 & 0.94 \\
         Accuracy & \multicolumn{2}{c}{0.97} \\
         \hline
    \end{tabular}
    \caption{Performance metrics, out-of-sample dataset. Target variable: sign of edges' impact}
    \label{tab:perf_oos}
\end{table}

\begin{figure}[t]
    \centering
    \includegraphics[width=0.8\linewidth]{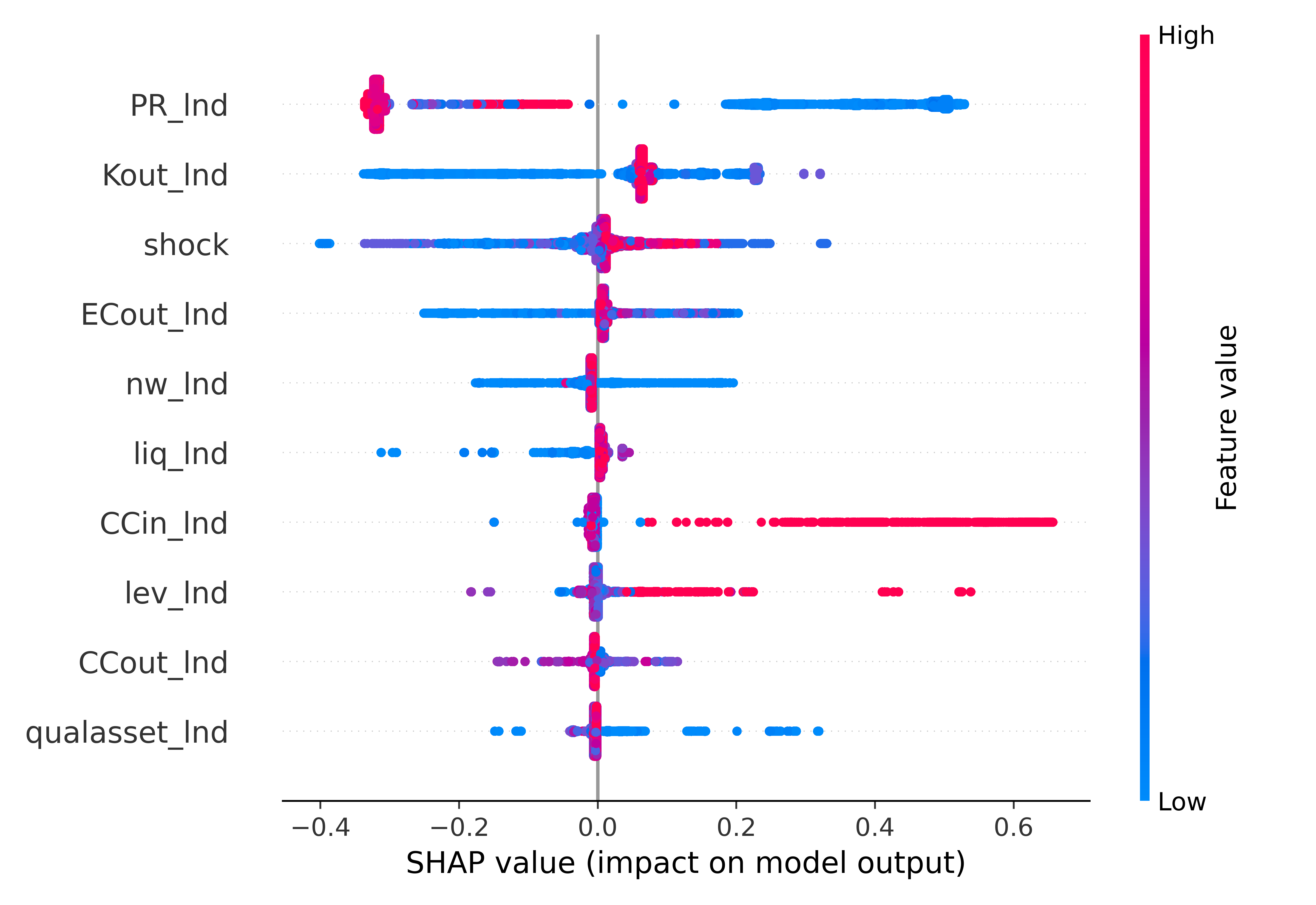}
    \caption{SHAP values, 10 most important features. Target variable: sign of edges’ impact}
    \label{fig:shap_sign}
\end{figure}

Figure \ref{fig:shap_sign} shows the ten most important features in driving the predicted output (i.e., those with the highest mean absolute SHAP value) in descending order. It can be observed that the sign of the edges' impact is mostly related to the PageRank and out-degree of the lender. (PR\_lnd and Kout\_lnd, respectively). Most of the instances in which the PageRank of the lender is smaller (i.e., the points closer to the blue spectrum) have their predicted value increased when this feature is included in the explainer model (i.e., they are at the right-hand side of the central axis). As a predicted value equal to 1 corresponds to an edge with a non-positive impact, the SHAP analysis suggests that the PageRank of the lender is negatively related to the probability of the sign of the edge's impact being non-positive.


The SHAP partial dependent plots (Figure \ref{fig:pdp_sign}) show that the relationship between the predicted output and the lender's PageRank displays a tipping point behavior. If the PageRank of the lender is below a given critical value $PR_{C}\approx0.0005$, the probability of a loan extended by this lender having a non-positive impact is superior to 80\%.\footnote{The value of $PR_{C}$ does not change significantly when a single value of $\zeta$ is considered.} Otherwise, this probability drops to zero (Figure \ref{fig:pdp_sign}, panel (a)). Thus, edges whose origin are lenders with a PageRank above this critical value have a greater chance of having a positive impact on the systemic risk. On the other hand, the probability of the edge having a non-positive impact increases when the out-degree of the lender surpasses a given threshold value, as can be seen in Figure \ref{fig:pdp_sign}, panel (b).

\section{Concluding remarks}
\label{sec:concl}
In this paper, we assessed how marginal variations in interconnectedness affect the stability of financial networks. To this aim, we presented a methodology to compute the impact of the edges in the stability of financial networks. This methodology was carried out in two sequential steps. First, we computed the SR of the financial network through the \textit{differential DebtRank} approach. Then we assessed how this measure of SR is impacted by the removal of the edge. The  percentage difference in the SR of the whole network caused by the removal of that edge is its impact. 

We applied this framework to a thorough Brazilian dataset to compute the impact of bank-firm edges in the stability of the financial network. This computation is performed considering different levels of the initial shock. Two main findings emerged from this preliminary analysis. First, there is a great heterogeneity concerning the impact of edges on the SR of the financial network. Some edges impact the stability of the financial network much more than others. Second, despite the impact of the edge on the SR is positive in most of the cases, the fraction of edges whose impact on the SR is null or negative increases with the level of the initial shock.

In light of these results, we performed two prediction exercises using ML techniques. First, we tried to predict the \textit{criticality} of the edges: an edge is deemed critical if it is an outlier according to the interquartile range (IQR) approach, considering its impact. Thus, a critical edge has a much greater impact on the SR of the financial network than the others. Second, we attempted to forecast the sign of the edges' impact. In both cases, topological and financial features of the edges' nodes (the origin and the destination node) were employed as predictive variables.

Concerning the critical status of the edges, the model proved to have high predictive power, showing a level of accuracy of over 90\%. A posterior interpretative analysis carried out through Shapley values showed the main predictor of the critical status of bank-firm edges is the PageRank of the borrower --- that is, the firm. Edges whose destination node has a large PageRank have a higher probability of being critical. The level of the initial shock and the eigenvector centrality of the lender are other important drivers of the edge's criticality. With respect to the second forecasting exercise, we found the sign of the edge's impact is mainly related to the PageRank of its origin node -- that is, the lender. The relationship between the predicted output and the lender’s PageRank displays a tipping point behavior. If the PageRank of the lender is below a given critical value of around 0.0005, the probability that a loan granted by this lender will have a negative impact is very large (around 80\%). Otherwise, this probability is close to zero.

This study contributes to the literature on the relationship between interconnectedness and systemic risk. Our results show that the impact of marginal changes in interconnectedness by removing or adding edges on the systemic risk is heterogeneous in magnitude and sign. In addition, we show which features -- related to the origin or the destination node -- identify an edge as critical or having a positive impact on the SR. Our results are also useful for policy-making purposes. They help financial regulators in the identification of financial transactions that will have the greatest impact on systemic risk and should therefore be closely monitored. Specifically, our results suggest that loans between firms and banks with a high PageRank should be discouraged, as such loans have a higher probability of having a positive and significant impact on the SR. Finally, by showing that topological features such as the PageRank are the most important drivers of the bank-firm edge's impact on SR, we corroborate the findings of other studies \citep{alexandre2021drivers,ghan2018,kuzubas2014,jaramillo2014}, according to which topological variables are at least as important as financial variables in driving systemic risk.

\newpage
\begin{appendices}
\section{Potential explanatory variables}
\setcounter{table}{0}
\renewcommand{\thetable}{A\arabic{table}}
\label{sec:app}

\begin{table}[H]
    \centering
    \captionsetup*{skip=0pt}
    \caption{List of potential explanatory variables}
    \begin{tabular}{llr}
    \hline
         Type & Description of the variable & Acronym \\
         \hline
         \multirow{8}[2]{40mm}{Lenders' F \& S variables} & Net worth & nw\_lnd\\ 
          & Return over equity & roe\_lnd\\
          & Leverage & lev\_lnd \\
          & Liquidity & liq\_lnd \\
          & Provisions-to-loans ratio & qualasset\_lnd \\
          & Dummy variable for private banks & private\_lnd \\
          & Dummy variable for foreign banks & foreign\_lnd \\
          & Dummy variable for credit unions & cred\_union\_lnd \\
        \hline
        \multirow{13}[2]{40mm}{Borrowers' F \& S variables} & Net worth & nw\_brw\\ 
          & Return over equity & roe\_brw\\
          & Leverage & lev\_brw \\
          & Bank debt-to-total debt ratio & bnkdebt\_brw \\
          & Market debt-to-total debt ratio & mktdebt\_brw \\
          & ROF debt-to-total debt ratio & rofdebt\_brw \\
          & Number of employees & qtt\_emp\_brw \\
          & Average wage & avg\_wage\_brw \\
          & Share of female employees & share\_women\_brw \\
          & Age of the firm & age\_brw \\
          & Short-term external debt ratio & extdebtst\_brw \\
          & Dummy for Limited Society & ltd\_brw \\
          & Dummies for the economic sector & sector\_brw\_$k^{*}$\\
        \hline
        \multirow{9}[2]{40mm}{Topological variables$^{**}$} & In-degree & Kin\\
        & Out-degree & Kout \\
        & Core number & KC \\
        & Closeness centrality (incoming links) & CCin \\
        & Closeness centrality (outgoing links) & CCout \\
        & Betweenness centrality & B \\
        & Eigenvector centrality (incoming links) & ECin \\
        & Eigenvector centrality (outgoing links) & ECout \\
        & PageRank & PR \\
        \hline
        \multicolumn{3}{l}{\footnotesize (*): $k$ refers to the economic sector, represented by letters from A to S.}\\
         \multicolumn{3}{l}{\footnotesize (**): The suffix "\_lnd" (for lenders) or "\_brw" (for borrowers) will be added to the acronym.}
    \end{tabular}
    \label{tab:vars}
\end{table}

\section{SHAP partial dependence plots}
\setcounter{figure}{0}
\renewcommand{\thefigure}{B\arabic{figure}}
\label{sec:spdp}

\begin{figure}[H]
\centering
\begin{subfigure}{.45\textwidth}
  \centering
  \includegraphics[width=1.0\linewidth]{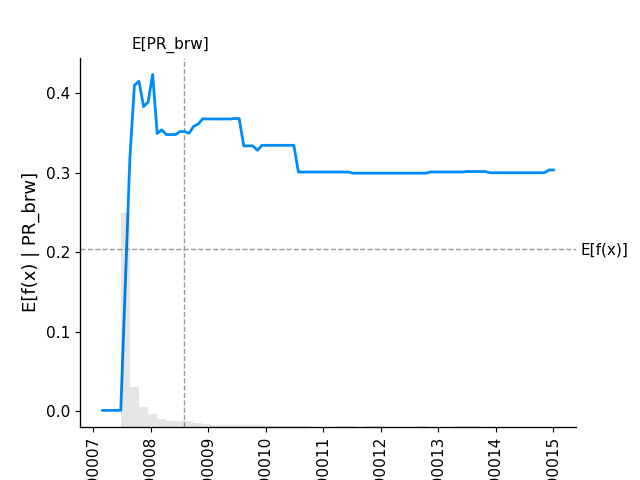}
  \caption{}
\end{subfigure}%
\begin{subfigure}{.45\textwidth}
  \centering
  \includegraphics[width=1.0\linewidth]{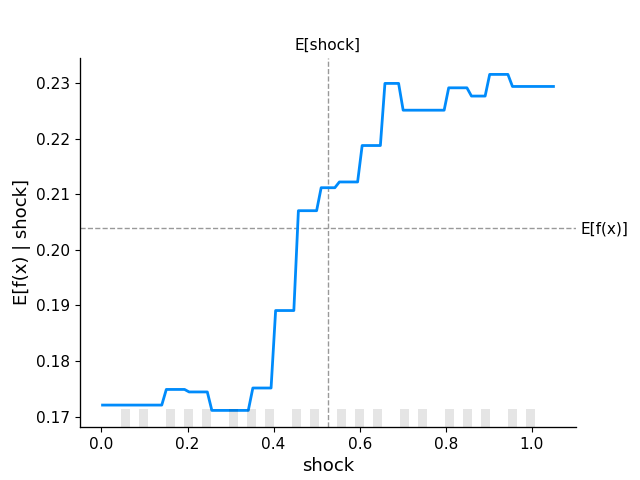}
  \caption{}
\end{subfigure}
\begin{subfigure}{.45\textwidth}
  \centering
  \includegraphics[width=1.0\linewidth]{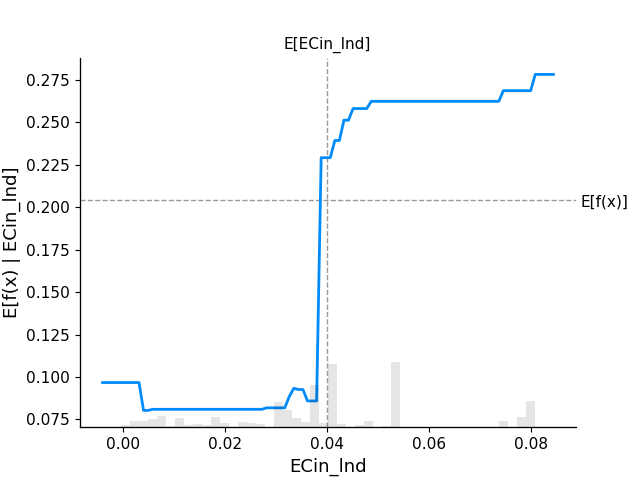}
  \caption{}
\end{subfigure}%
\begin{subfigure}{.45\textwidth}
  \centering
  \includegraphics[width=1.0\linewidth]{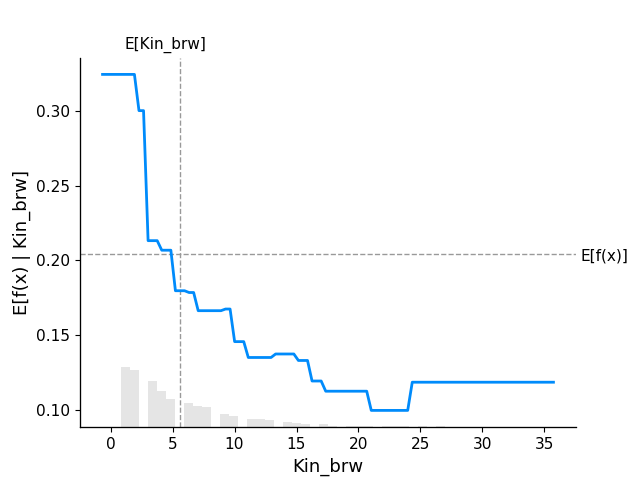}
  \caption{}
\end{subfigure}
\caption{SHAP partial dependence plots, most important features. Target variable: edges' critical status}
\label{fig:pdp_otl}
\end{figure}

\begin{figure}[H]
\centering
\begin{subfigure}{.45\textwidth}
  \centering
  \includegraphics[width=1.0\linewidth]{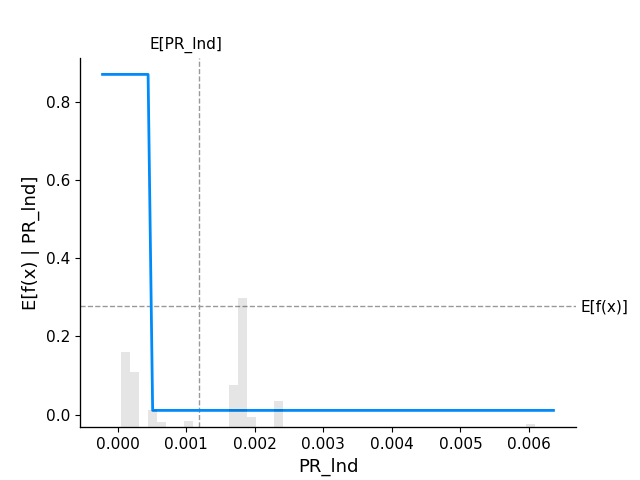}
  \caption{}
\end{subfigure}%
\begin{subfigure}{.45\textwidth}
  \centering
  \includegraphics[width=1.0\linewidth]{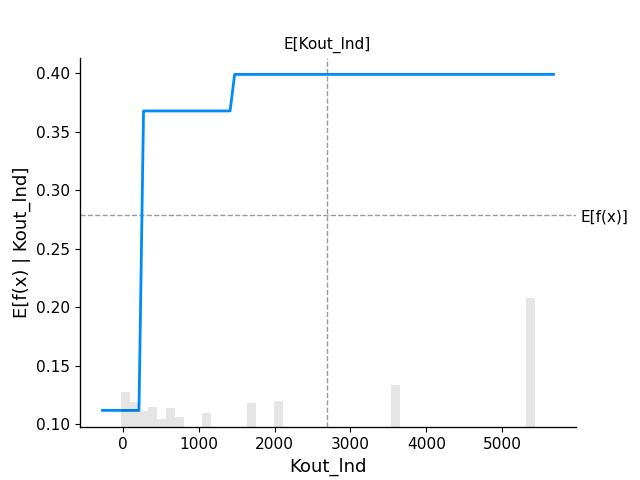}
  \caption{}
\end{subfigure}
\begin{subfigure}{.45\textwidth}
  \centering
  \includegraphics[width=1.0\linewidth]{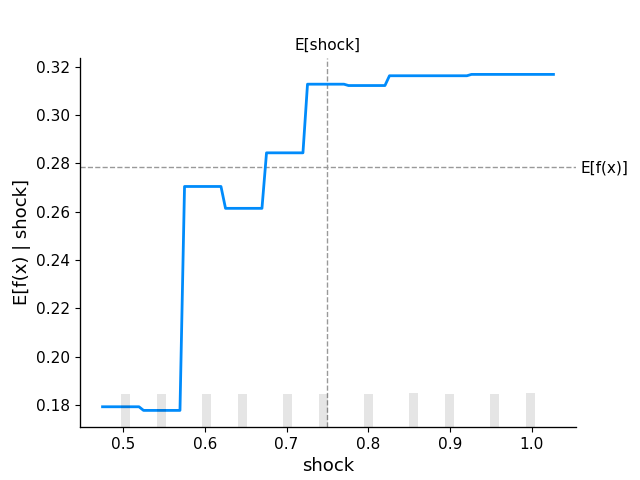}
  \caption{}
\end{subfigure}%
\begin{subfigure}{.45\textwidth}
  \centering
  \includegraphics[width=1.0\linewidth]{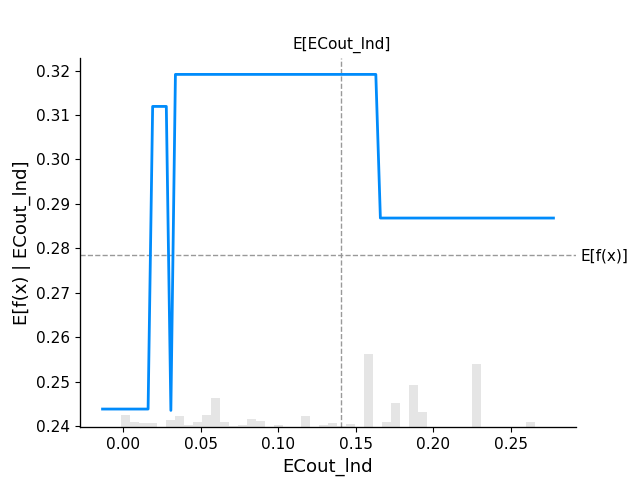}
  \caption{}
\end{subfigure}
\caption{SHAP partial dependence plots, most important features. Target variable: sign of edges' impact}
\label{fig:pdp_sign}
\end{figure}

\end{appendices}

\newpage
\singlespacing
\bibliographystyle{ecta}
\bibliography{Bibliography/mybibfile}

\end{document}